\documentclass[11pt,twoside]{article}

\usepackage{asp2006}
\usepackage{epsf}
\usepackage{lscape}

\newcommand{\msun}{M_{\odot}}

\begin{document}
\title{From Massive Cores to Massive Stars}   
\author{Mark R. Krumholz}   
\affil{Department of Astrophysical Sciences, Princeton University, Princeton, NJ 08544-1001 USA}    

\begin{abstract} 
The similarity between the mass and spatial distributions of
pre-stellar gas cores in star-forming clouds and young stars in
clusters provides strong circumstantial evidence that these gas cores
are the direct progenitors of individual stars. Here I describe a
physical model for the evolution of massive cores into stars, starting
with the intial phases of collapse and fragmentation, through disk
formation and fragmentation, the later phases of stellar feedback, and
finally interaction of the newly formed stars with their
environments. This model shows that a direct mapping from cores to
stars is the natural physical outcome of massive core evolution, and
thereby allows us to explain many of the properties of young star
clusters as direct imprints of their gas-phase progenitors.
\end{abstract}



\section{Introduction}

Massive stars form in regions of extremely high column density, hidden
behind hundreds of magnitudes of visual extinction. As radio and
submillimeter interferometers have matured over the past decade, they have
revealed the properties of these regions with ever higher detail, to
the point where today it is not entirely unreasonable to speak of
observationally determined ``initial conditions'' for the problem of
star formation. One of the most striking results of this exploration
has been the extent to which the properties of young star clusters
are directly mirrored in the conditions found in pre-stellar molecular gas. 

As the most basic level, young clusters and the molecular clumps from which they form have similar bulk
properties, such as column density ($\sim 1$ g cm$^{-2}$), size
($\sim 1$ pc), and velocity dispersion (a
few km s$^{-1}$) \citep{mckee03}. More interestingly, the
dense cores within these clumps also mirror the properties of
stars. Cores are bound, centrally-condensed objects with
characteristic sizes $\sim 0.1$ pc or smaller and masses comparable to
those of individual stars \citep{sridharan05,beuther05b}. Observations in many regions with a variety of techniques find that the core and star mass functions have the same shape,
differing only in that cores are a factor of
$2-4$ more massive \citep[e.g.][and J. Alves, this volume]{motte98, johnstone01,
reid06b, alves07}. Moreover, cores appear to be mass-segregated: cores with masses greater than a few $\msun$ are found only in the centers of their parent clumps, but the core mass function is
otherwise independent of position \citep{elmegreen01, stanke06}. This is remarkably similar to the
pattern in young clusters, where there is no segregation for stars smaller than a few $\msun$, but more massive stars are almost exclusively in cluster centers \citep{hillenbrand98, huff06}.

The coincidence in both the mass and spatial distributions of cores
and stars makes a strong circumstantial case that young stars'
properties might be direct imprints of core properties. (The somewhat
higher core masses are to be expected, since outflows should
prevent $\sim 50\%$ of a core's mass from reaching a star --
\citealt{matzner00}.) However, we
cannot make such an inference without a theoretical understanding of
how cores might evolve into stars. The goal of this paper is to
summarize work in the last few years that has focused on the most
problematic part of this process, understanding the evolution of
massive cores. In the following sections I sketch a model for the
evolution of these objects, beginning with observed core properties and using
numeric and analytic arguments to understand how they collapse into
stars.

\section{Initial Collapse and Framgentation}

The first phase of evolution for a massive core begins when it starts to collapse but has not yet formed any stars. One might initally suspect that there is no plausible way for a
massive core to collapse to a single star or a small-multiple system,
since the Jeans mass in cold molecular clumps is only $\sim
\msun$. If massive cores do truly fragment into Jeans mass-sized
objects once their collapse begins, then there can be no direct mapping from cores to stars. Such
behavior is exactly what purely hydrodynamic numerical simulations
find: as objects collapse and their density rises, the Jeans mass
falls, and the objects break into smaller and smaller pieces that
always have masses comparable to the Jeans mass at their current
density. The fragmentation process ceases only when the assumed
equation of state stiffens, so massive cores generate numerous small
stars \citep[e.g.][]{bate05,dobbs05}.

\begin{figure}
\plotone{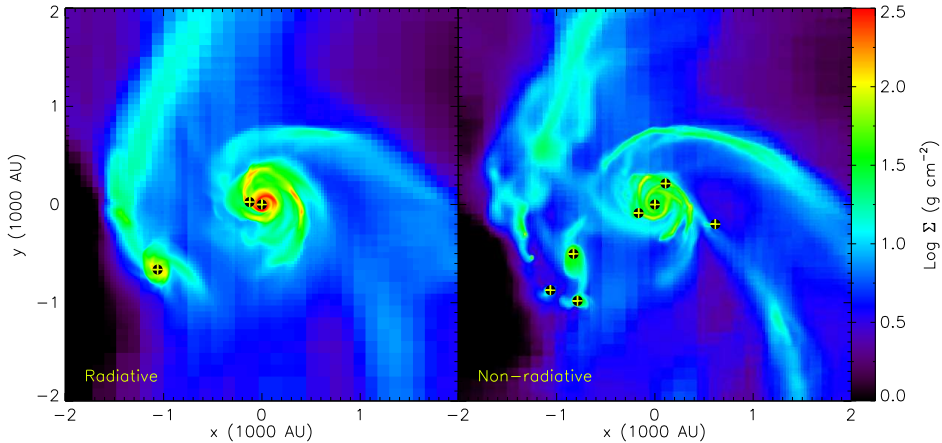}
\caption{
\label{rtcomparison}
Column density in two simulations of the collapse and fragmentation of
a $100$ $\msun$ core. The left and right panels show two runs from
\citet[runs 100A and 100ISO]{krumholz07a} at 20 kyr of evolution; one
uses radiative transfer and one does not, but otherwise the
simulations have identical initial conditions and resolution. The plus
signs indicate the positions of stars. Note the significantly greater
number of stars, with several more condensations on the verge of
collapse to stars, in the non-radiative run.
}
\end{figure}

However, while a purely hydrodynamic approach to fragmentation is
analytically very simple and numerically very cheap, it neglects the
important effect of radiation feedback from embedded, forming
protostars. \citet{krumholz06b} points out that, even before embedded
stars begin nuclear burning, just the gravitational energy released as
gas accretes onto them can significantly heat the surrounding gas,
raising the Jeans mass and suppressing
fragmentation. \citet{krumholz07a} follow up this point by
simulating the collapse of turbulent, massive cores whose initial
masses, sizes, central concentrations, and levels of turbulence
are chosen to match those of observed cores
(as seen e.g.\ by \citealt{sridharan05} and \citealt{beuther05b}). The
simulations combine a protostellar evolution model with a new
adaptive mesh refinement gravity-radiation-hydrodynamics algorithm
\citep{krumholz04,krumholz07b} to model accurately the effects of
radiative heating. The simulations show that radiative heating
strongly suppresses fragmentation of massive cores, allowing the great
majority of the mass in a core to collapse into one or two stars. In
contrast, a control run omitting radiative heating qualitatively
reproduces the earlier hydrodynamic result that massive cores collapse
into dozens of small fragments. Figure \ref{rtcomparison} illustrates
the difference made by the inclusion of radiative heating.

The most important point to take from this is that, when more detailed
physics than simple hydodynamics is included, simulations and analytic
arguments both show that the observed massive cores are unlikely to
fragment into many pieces. Thus, a model in which there is a direct
mapping from the masses and positions of massive cores to those of
massive stars passes its first test: the cores will collapse largely
monolithically, rather than fragmenting to small stars. A secondary
point is that we must be careful about drawing conclusions based on simulations with very simple physics. Models that do
not include radiative transfer produce qualitatively different results
from those that do.

\section{Disk and Binary Formation}

\begin{figure}
\plotone{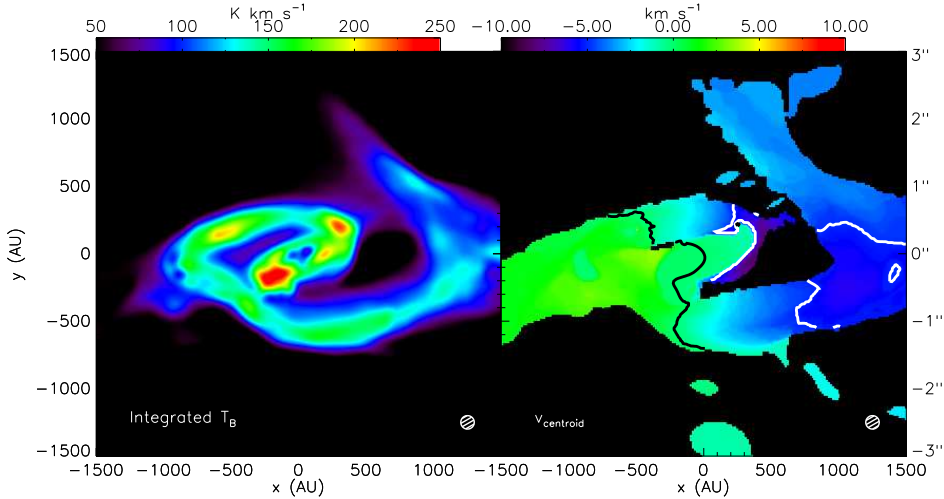}
\caption{
\label{diskobs}
The velocity-integrated brightness temperature (\textit{left panel})
and centroid velocity (\textit{right panel}) in a simulated $0."1$ resolution ALMA
observation of a massive protostellar disk 500 pc away in the CH$_3$CN 220.7472
GHz line. The disk is from the simulation illustrated in the
left panel of Figure \ref{rtcomparison} at an evolution time of 27
kyr, when the central star is $8.3$ $\msun$ in mass. The white contour
in the right panel corresponds to $-5$ km s$^{-1}$, and the black
contour shows $0$ km s$^{-1}$, measured relative to the velocity of
the central star. The disk is systematically offset to negative
velocities relative to the star, so while there is a large region with
velocity $<-5$ km s$^{-1}$, there are no pixels with
velocities $> 5$ km s$^{-1}$. Black pixels correspond to locations
where ALMA would not detect the line at $>3\sigma$ confidence.
}
\end{figure}

Since collapsing turbulent cores have non-zero angular momentum, they
naturally form protostellar disks. Observationally, these are a
potential signpost of the star formation process. Both the simulations
described above and analytic models \citep{kratter06} find that
the disks formed by massive cores are likely to be strongly
gravitationally unstable. This instability causes
the disks to develop large-amplitude $m=1$ spiral modes, and potentially
even fragment to form companions to the primary star (although a
majority of the mass still goes into the primary, not the
fragments). \citet{krumholz07d} shows that the strong $m=1$ spiral
structure present in such unstable disks should be observable with
next-generation telescopes such as ALMA and the EVLA, and that a
systematic offset between the disk's ``zero'' velocity and the central
star's velocity produced by the instability might also be
detectable. The detection of a disk with these signatures inside a
massive core would be strong evidence in favor of the model that
massive cores collapse to form massive stars. Figure \ref{diskobs}
shows a simulated ALMA observation of such a disk, computed using the
technique of \citet{krumholz07d}.

Disks around massive stars are also important for their role
in forming companions to massive stars. Most massive stars have
close companions \citep{lada06}, and gravitational instability is
massive protostellar disks provides a natural explanation for
this because even radiatively heated massive disks suffer some fragmentation
(though vastly less than if radiation is omitted). These fragments initially form with masses
$\sim \msun$ at distances $> 100$ AU from the primary
\citep{kratter06,krumholz07a}, but they subsequently migrate
inward to separations $<10$ AU as the disk accretes. The final separations of these disk-formed companions from the primary, and whether some of them merge with it, has not yet been determined.

An interesting fate awaits migrating stars that get close
to the primary but do not merge with it. Massive protostars go through
a phase of deuterium shell burning, during which their radii swell
to tenths of an AU in size. \citet{krumholz07c} point out that this
may lead a primary to overflow its Roche lobe and transfer mass onto
close companions. Such mass transfer is almost
always unstable, terminating only once the mass ratio of the system
reaches unity. This provides a natural mechanism for the origin of the heretofore unexplained massive ``twins'', binaries consisting of two massive stars with a mass
ratio of almost exactly unity \citep[e.g.][]{pinsonneault06}.

\section{Radiation Pressure Feedback}

A direct core to star mapping is possible only if most of a
core's mass be able to accrete onto the massive star forming within
it. However, spherically
symmetric calculations indicate that the huge radiation output
output of massive stars should exert a force stronger than gravity on
dust grains suspended in the gas around them for stars of $\sim 20$ $\msun$ or larger
\citep{larson71, yorke77, wolfire87}. How massive stars can form despite this radiation barrier is a classic problem in astrophysics. Fortunately, rotation and the formation of a
disk can significantly mitigate this effect, because gas in a disk self-shields against the radiation due to high optical depths \citep{nakano89, nakano95, jijina96}, while at the same time a disk collimates the
radiation field and beams it away preferentially in the polar direction,
thereby reducing the radation force felt by gas in the equatorial
plane \citep{yorke02}. However, even with these effects it is not
entirely clear that stars can grow to arbitrary masses by accretion.

Two additional effects may help. First, radiation hydrodynamic
simulations show that the first effect of radiation pressure is that
massive stars blow radiation bubbles above and below accretion disks
\citep{yorke02,krumholz05d}. However, in three dimensions these bubbles
do not halt accretion, because gas that reaches the bubble wall
flows along the wall onto the accretion disk. Bubbles may also
collapse due to Rayleigh-Taylor
instability, allowing accretion to continue through optically thick
channels while radiation escapes through optically thin regions around
them \citep{krumholz05d}. If magnetic fields are present and sufficiently strong, this effect will be enhanced by photon bubble instability \citep{turner07}, which arranges the gas into dense lumps separated by low-density gaps through which radiation leaks, effectively reducing the radiation pressure force experienced by the bulk of the gas.

Second, protostellar
outflows provide a third escape valve for radiation. Massive
protostars appear to generate hydromagnetic outflows just like low
mass stars, with the difference that for massive stars the outflow
cavities are largely dust-free because the base of the outflow is
close enough to the star for the dust within it to have been destroyed
by sublimation. Such outflow cavities therefore present optically thin
channels through which radiation can leak out of the optically thick
cores. Radiative transfer calculations show that this can lead to
order-of-magnitude reductions in the radiation force on accreting gas
near the equatorial plane, again allowing accretion to continue where
it might otherwise have been halted \citep{krumholz05a}.

While a definitive numerical simulation including the effects of
radiation forces, magnetic fields, and protostellar outflows in three
dimensions has not yet been done, and is probably at best barely within
the capabilities of present-day supercomputers, it seems clear that
each of these effects will help massive stars form by accretion. Thus,
we can tentatively say that there is no barrier to most of the gas in
a protostellar core accreting onto a massive star. Conversely, however, simulations of massive star formation that do not include radiation force effects are simply ignoring
this problem entirely. Preliminary simulations indicate that
models of star formation that depend on Bondi-Hoyle accretion are
likely to fail once radiation pressure is included \citep{edgar04}, so models of this sort must be viewed with caution.

\section{Competitive Accretion}

Thus far we have seen that massive protostellar cores will not
fragment strongly, what fragmentation they do show is consistent with
the observed multiplicity properties of massive stars, and that
radiation pressure will not prevent most of the mass in a core from
accreting onto a star. There remains, however, one more way in which a
direct mapping from cores to stars could fail: if, once stars accrete
their parent cores, they were subsequently to accrete a great deal
more mass, then there would be no direct relationship between core and
stellar masses. This process of accretion onto stars from gas that was
not originally part of a bound core is known as competitive accretion
\citep{bonnell01a,bonnell01b}.

Purely hydrodynamic simulations of star cluster formation show that
this process is the dominant mechanism by which stars gain mass. In
effect, all stars are born at masses of order the Jeans mass, but some
of them fall to the center of the collapsing gas cloud, and the deep
potential well then channels gas to them. They subsequently accrete
this gas and grow in mass, producing a full range of initial masses
\citep{bonnell04, bonnell06c}.

However, direct observational estimates of the rate of competitive accretion
generally find that it is too small to make a significant contribution
to final stellar masses \citep{andre07}.
\citet{krumholz05e,krumholz06a} point out that competitive accretion
is possibly in simulations only because the simulated gas clumps are
in the process of global collapse, which creates deep potential wells
within which the gas is dense and non-turbulent and Bondi-Hoyle accretion is
rapid. These deep, dense, quiescent gas wells have not been observed, however. This is probably because clumps are not in a state of global collapse. Such a collapse necessarily converts order unity of the mass in a gas clump into stars and ends star formation in $1-2$ free-fall times \citep[e.g.][]{bonnell04}. However, there is strong evidence that the star formation process cannot be anywhere near that fast. The gas clump from which the ONC formed likely had a density $\sim 10^5$ cm$^{-3}$ \citep{elmegreen00}, implying a free-fall time of $0.1-0.2$ Myr, but the estimated ages of the stars in Orion point to a formation process lasting $1-3$ Myr, implying a minimum formation time scale of 5 free-fall times even if one assumes the fastest plausible formation time, with $\sim 15$ free-fall times being more likely \citep{tan06a}. Furthermore, the total galactic mass of infrared dark clouds, which have densities $\sim 10^3$ cm$^{-3}$, is such that, if they were collapsing to form stars on a free-fall time scale, the total galactic star formation rate would have to be $\sim 100$ times higher than its observed value. The same conclusion holds for dense gas clumps traced by HCN($1\rightarrow0$) emission, which have densities $\sim 10^4-10^5$ cm$^{-3}$ \citep{krumholz06c}. This implies that, even in regions with mean densities of $10^5$ cm$^{-3}$ star formation cannot be anywhere near as rapid as would be required for competitive accretion to occur.

The final piece of evidence against competitive accretion comes from simulations that include more detailed physics. \citet{li06b} and \citet{nakamura07} simulate the formation of a star cluster using magnetohydrodynamics rather than simple hydrodynamics, and including the effects of outflows driven by the protostars forming within the cluster. They find that the protostellar outflows drive turbulent motions, preventing global collapse and ensuring that conditions are too turbulent for competitive accretion processes to alter the masses of protostars significantly after they have consumed their parent cores. Moreover, these simulations, unlike ones where competitive accretion occurs, produce star formation rates in good agreement with observations. We can therefore tentatively conclude that the core accretion hypothesis passes this final test: once stars have accreted their parent cores, they will not gain much additional mass from the gas to which they were not bound at birth. A direct core to star mapping will survive.

\section{Summary and Conclusion}

For the past decade observations have increasingly pointed to an intimate link between young stars and the dense stellar-mass gas clouds known as pre-stellar cores. Cores and stars have very similar mass and spatial distributions, so it is tempting to explain the properties of young star clusters as imprinted at birth. However, this hypothesis requires that cores map directly onto stars. Here I provide a physical model for such a mapping. When massive cores first collapse, they do not fragment strongly because the first stars that form within them heat the gas and suppress fragmentation. As a result, massive cores collapse to only a few stars. Most of the fragmentation that does happen occurs in unstable self-gravitating disks, which should be directly observable. Disk fragmentation ensures that massive stars will essentially always have companions, and some of these companions will turn into twins of the primary star. As the massive star grows, it will begin to generate a huge radiation force opposing accretion. However, a combination of instabilities and the leakage of radiation through protostellar outflow cavities allows the radiation to escape and accretion to continue unimpeded to high masses. Finally, once stars have accreted their parent cores, they will be unable to gain additional mass that was not originally part of the core. Thus, the hypothesis that stars come directly from cores, with a one-to-one mapping of core mass to star mass, is in agreement with a physical model for how massive cores evolve. The best explanation for many properties of young star clusters appears to be that they are set at birth, when the cluster is still a dark cloud.

\acknowledgements My work on massive stars is done in collaboration with R. Klein, K. Kratter, C. Matzner, C. McKee, J. Tan, and T. Thompson. Support for this work was
provided by NASA through Hubble Fellowship grant \#HSF-HF-01186
awarded by the Space Telescope Science Institute, which is operated by
the Association of Universities for Research in Astronomy, Inc., for
NASA, under contract NAS 5-26555. The simulations described were made
possible by grants of high performance computing resources from
the Arctic Region Supercomputing Center; the NSF San Diego
Supercomputer Center through NPACI program grant UCB267; the National
Energy Research Scientific Computing Center, which is supported by the
Office of Science of the U.S. Department of Energy under Contract
No. DE-AC03-76SF00098, through ERCAP grant 80325; and the US
Department of Energy at the Lawrence Livermore National Laboratory
under contract W-7405-Eng-48.




\end{document}